\documentclass[twocolumn,nofootinbib]{revtex4}

\usepackage{amsmath}
\usepackage{amsfonts,amssymb,amsthm, bbm,braket}
\usepackage{graphicx}   
\usepackage{subfigure}  
\usepackage{amsbsy} 
\usepackage[bold]{hhtensor} 
\usepackage{natbib}
\usepackage{algpseudocode}

\usepackage{multirow}

\newcommand\Tr{\mathrm{Tr}}

\usepackage{hyperref} 

\hypersetup{
  colorlinks   = true, 
  urlcolor     = blue, 
  linkcolor    = blue, 
  citecolor   =  red 
}

\begin{document}
\title{Self-guided quantum tomography}

\author{Christopher Ferrie}
\affiliation{
Center for Quantum Information and Control,
University of New Mexico,
Albuquerque, New Mexico, 87131-0001}

\date{\today}


\begin{abstract}
We introduce a self-learning tomographic technique in which the experiment guides itself to an estimate of its own state.  Self-guided quantum tomography (SGQT) uses measurements to directly test hypotheses in an iterative algorithm which converges to the true state.   We demonstrate through simulation on many qubits that SGQT is a more efficient and robust alternative to the usual paradigm of taking a large amount of informationally complete data and solving the inverse problem of post-processed state estimation.
\end{abstract}


\maketitle

The act of inferring a quantum mechanical description of a physical system---assigning it a \emph{quantum state}---is referred to as tomography.  Tomography is a required, and now routine, task for designing, testing and tuning qubits---the building blocks of a quantum information processing device \cite{nielsen2010quantum}.  However, in a grand irony, the exact same exponential scaling that gives a quantum information processing device its power also limits our ability to characterize it.  

That tomography is a problem exponentially hard in the number of qubits has lead to many proposals for efficient learning within restricted subsets of quantum states \cite{Cramer2010Efficient,Toth2010Permutationally}.  On the other hand, if we expect to have prepared a specific target state, efficient protocols exist to estimate the fidelity to this state \cite{flammia2011direct, da2011practical}.  These protocols are \emph{direct} in the sense that few measurements are required to provide an estimate of the fidelity to the target rather than first reconstructing the state then calculating the fidelity.

The proposal presented here is direct in the same sense as \cite{flammia2011direct, da2011practical}, but converges to the state itself.  The algorithm is iterative---from directly estimating a distance measure to the underlying state, the experiment guides itself to a description of its own state: \emph{self-guided quantum tomography} (SGQT).  The distance measure $m$ is arbitrary, in the sense that any measure will work.  However, the more rapidly the experiment can provide an estimate for $m$, the more rapidly SGQT will converge, such that if $m$ can be estimated \emph{efficiently}, then SGQT will be efficient.

Before describing exactly what SGQT is, we first state what it is not by reviewing the problem of tomography.  There is some true state $\rho$ which generates data, a list of measurement outcomes corresponding to effects $D=\{E_0,E_1,\ldots\}$.  The probability to observe this data is given by the Born rule
\begin{equation}
\Pr(D|\rho) = \prod_k \Tr(\rho E_k).
\end{equation} 
The prevailing method is to solve the inverse problem of identifying an accurate estimate, $\sigma$, of $\rho$ given a sample data set drawn from this distribution.  Here the approach is quite different.  We begin with a distance measure on states $m(\rho,\sigma)$. The only requirement is that this measure can be estimated from experiment, such that we have access to
\begin{equation}
f(\sigma) = \langle m(\rho,\sigma) \rangle.
\end{equation}
This quantity fluctuates from noise which can come from a variety of sources but is always present due to the fundamental statistical nature of quantum mechanics---also known as \emph{shot-noise}.

Here we will provide an algorithm to iteratively propose new states $\sigma$ such that we converge to $\rho$ only via estimates of $f(\sigma)$.  The core of the algorithm is a stochastic optimization technique known as simultaneous perturbation stochastic approximation (SPSA) \cite{spall1992multivariate}.

SPSA is an iterative optimization technique which uses only two (noisy) function calls per iteration to estimate the gradient.  In the context of state estimation, this means that SGQT requires only two proposal states $\sigma_{\pm}$ and experimental estimates $f(\sigma_{\pm})$ to provide an unbiased estimate of the gradient, which in turn provides the direction to the true state.  This is the key element which provides SGQT with its efficiency.  For one might expect that to estimate the gradient would require $O(2^{2n})$ proposal states, where $n$ is the number of qubits.  In the remainder, we detail the algorithm and demonstrate via numerical experiments the claimed efficiency and robustness of SGQT.

The steps of each iteration proceed as follows (suppose we are at iteration $k$): (1) Generate a random direction to search in defined by ${\triangle}_k$.  (2)
Calculate the estimated gradient in that direction,
\begin{equation}
g_k = \frac{f(\sigma_k+\beta_k{\triangle}_k)-f(\sigma_k-\beta_k{\triangle}_k)}{2\beta_k}{\triangle}_k.
\end{equation}
(3) Calculate the next iterate via
\begin{equation}
\sigma_{k+1} = \sigma_{k} + \alpha_k {g}_k. 
 \end{equation}
The functions $\alpha_k$ and $\beta_k$ control the convergence and are user defined, although they are usually specified in the forms
\begin{equation}
\alpha_k = \frac{a}{(k+1+ A)^s},\;\; \beta_k= \frac{b}{(k+1)^t},
\end{equation}
where $a,A, b,s$ and $t$ are chosen first roughly based on extensive numerical studies for many problems then tweaked based on numerical simulations for the problem at hand.  It is useful to note that much of the former task has been done and generally good choices are \cite{sadegh1996optimal} $s=0.602$ and $t=0.101$. These values indeed work for our problem but the asymptotically optimal \cite{spall1992multivariate} values $s=1$ and $t=1/6$ seems to perform well even early on.  The random direction $\triangle_k$ is arbitrary, up to some constraints.  Given some vector representation of the operators, we take the common choice $\triangle_k = \pm 1$ for each element of the vector and the sign randomly assigned by a fair coin toss.  For $a, A$ and $b$, the optimal parameter are more problem dependent and here we have two different sets depending on whether we are demonstrating asymptotic performance or not---these values will be noted when the results are discussed later on.  Convergence results on SPSA \cite{spall1992multivariate,sadegh1996optimal} imply that the infidelity, for example, decreases at rate $O(1/k^{\gamma})$, where the exponent actually achieved is highly problem-dependent.  Asymptotic results, however, give a rate $\gamma \approx 1$ to first order.

\begin{figure}\centering
  \includegraphics[width=\columnwidth]{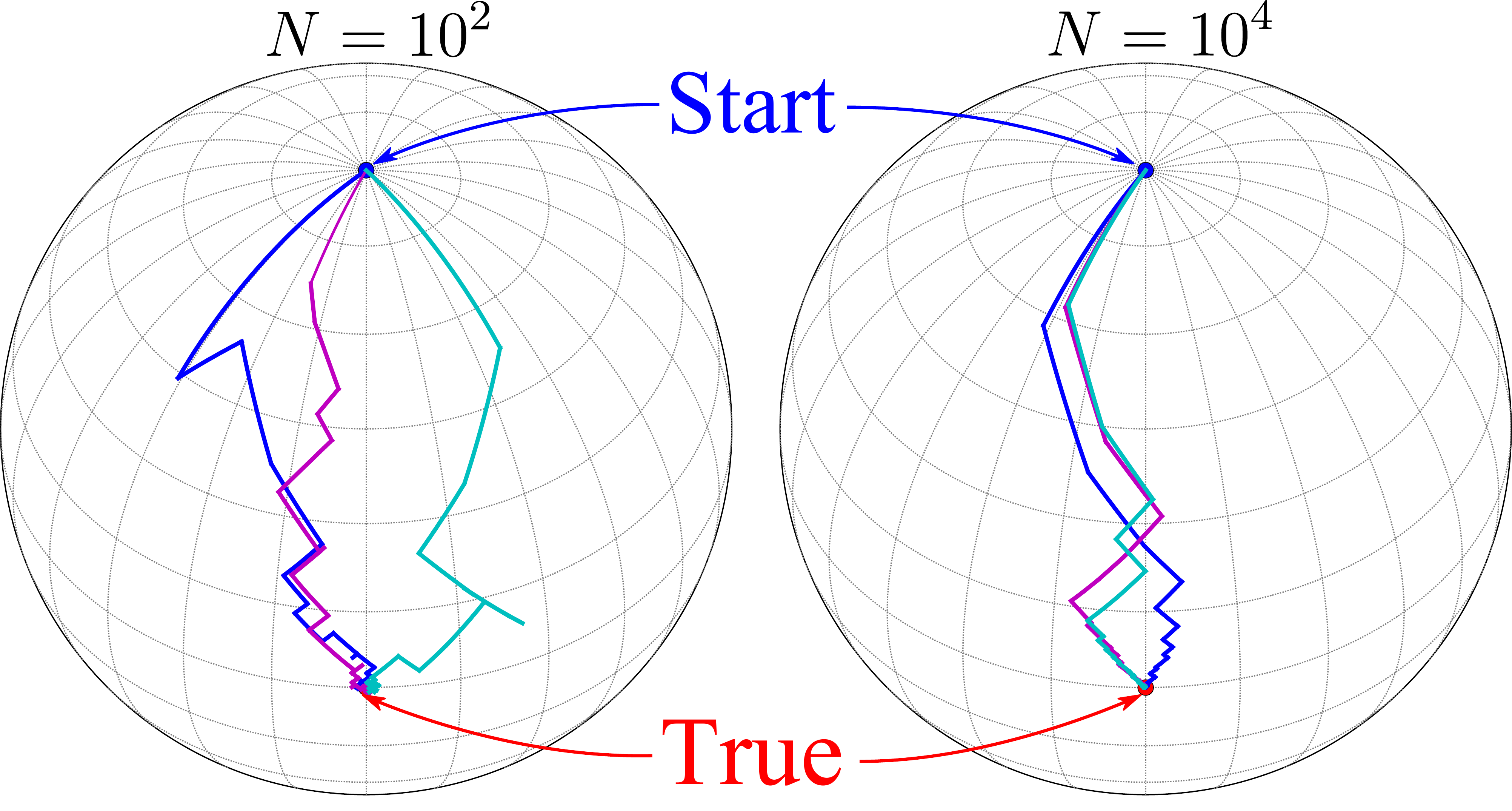}
  \caption{\label{fig:qubit_vis} Estimating the state of a qubit via SGQT and the fidelity metric.  On the left, each fidelity estimate was generated with $N = 10^2$ single shot experiments---on the right, $N=10^4$.  Each line represents one of three runs with $k = 10^3$ iterations.}
\end{figure}

We illustrate the iterations of the algorithm in Fig.~\ref{fig:qubit_vis} for the metric
$m(\rho,\sigma) = 1- F(\rho,\sigma)$ the infidelity between the two states.  For pure states this is equivalent to $m(\psi,\phi) = 1-|\langle \psi|\phi\rangle|^2$ and can be estimated by measuring in the basis containing $\ket\phi$.  That is, by counting the number of outcomes in the direction of $\ket\phi$, say $n(\phi_+)$, we can estimate
\begin{equation}\label{eq:fid_est}
m(\psi,\phi) \approx 1 - \frac{n(\phi_+)}{n(\phi_+)+n(\phi_-)},
\end{equation}
which fluctuates due to statistical noise.  In Fig.~\ref{fig:qubit_vis}, we see this manifest through the volatility of the path taken by the algorithm when $n(\phi_+)+n(\phi_-)\equiv N=10^2$ and $N=10^4$.  We might expect then that more experiments are needed to mitigate these fluctuations as we converge on the target true state.  We will see, however, that this intuition fails us.  That is, for a fixed number of iterations, the performance is roughly independent of the number of experiments.  This will demonstrate the superior efficiency of SGQT to converge well beyond what we might expect to be the ``noise floor''.

In our discussion, we will refer to the follow three algorithmic and experimental parameters: $N$, the number of experiments per estimate of $m$; $M$, the number of estimates of $m$ per iteration; and $k$, the number of iterations.  Thus, the total number of experiments is $N_{\rm tot}=N \cdot M \cdot k$.  For standard finite difference gradient estimation, we have $M = 2 d$, where $d$ is the real dimension of the state space.  For $n$ pure qubits $d = 2(2^{n} -1)$, which grows exponentially.  For SGQT, however, $M=2$ regardless of the dimension, thus we will restrict our attention to $N$ and $k$ with the understanding that $N_{\rm tot} = 2 Nk$. 

\begin{figure}[h]\centering
  \includegraphics[width=1\columnwidth]{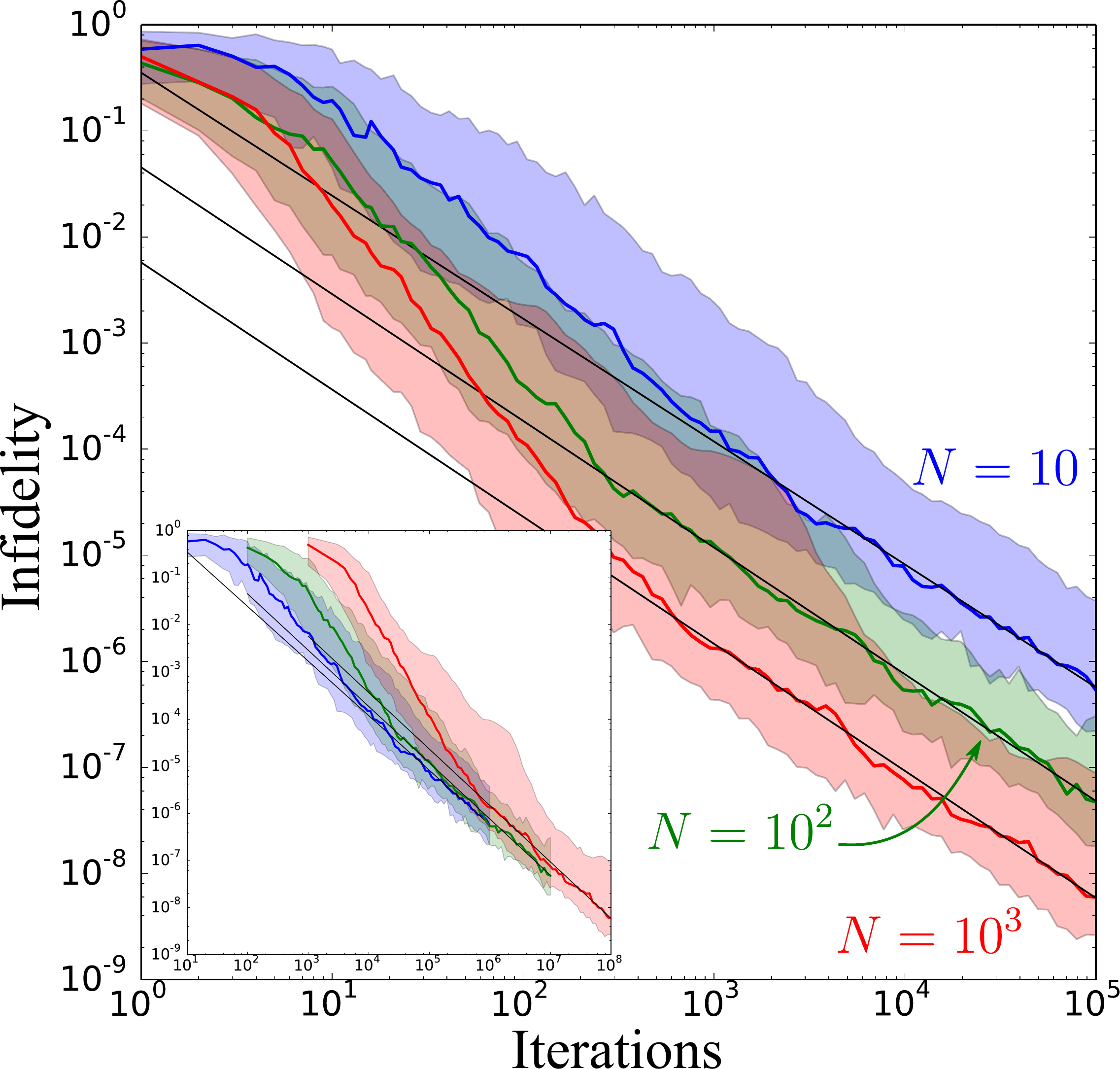}
  \caption{\label{fig:qubit_v_k} The infidelity vs the number of iterations $k$ achieved by SGQT.  Each line is the median performance of SGQT over 100 randomly (according to Haar measure) chosen pure states.  The shaded regions represent the interquartile range of infidelities.  The inset shows the performance as function $N\cdot k$ by simply shifting the original lines by their corresponding $N$.  The SGQT parameters chosen for these simulation where $a=3$, $A=0$ and $b=0.1$.}
\end{figure}

We continue with the qubit example of Fig.~\ref{fig:qubit_vis} to determine how the performance of SGQT scales with $N$ and $k$ retaining the fidelity objective function in Eq.~\eqref{eq:fid_est}.  In Fig.~\ref{fig:qubit_v_k}, we plot the infidelity as a function $k$.  We find, independent of $N$, the asymptotic scaling of infidelity is $O(1/k^\gamma)$ with $\gamma\in(1.16,1.20)$.  This is slightly better than what we would expect from the asymptotic rate of $O(1/k)$.

The inset in Fig.~\ref{fig:qubit_v_k} shows that in the asymptotic regime, the performance is roughly independent $N$, but also shows that, in terms of the total number of experiments, fewer repetitions per measurement setting is better initially.  That is, contrary to what we might expect, it is not necessary to increase the number of experimental repetitions to increase the accuracy of the estimated fidelity.  This false intuition would, however, hold true if we were to use an optimization algorithm (such as a standard gradient descent) which does not take account of the stochasticity in estimating the fidelity.

\begin{figure}\centering
  \includegraphics[width=1\columnwidth]{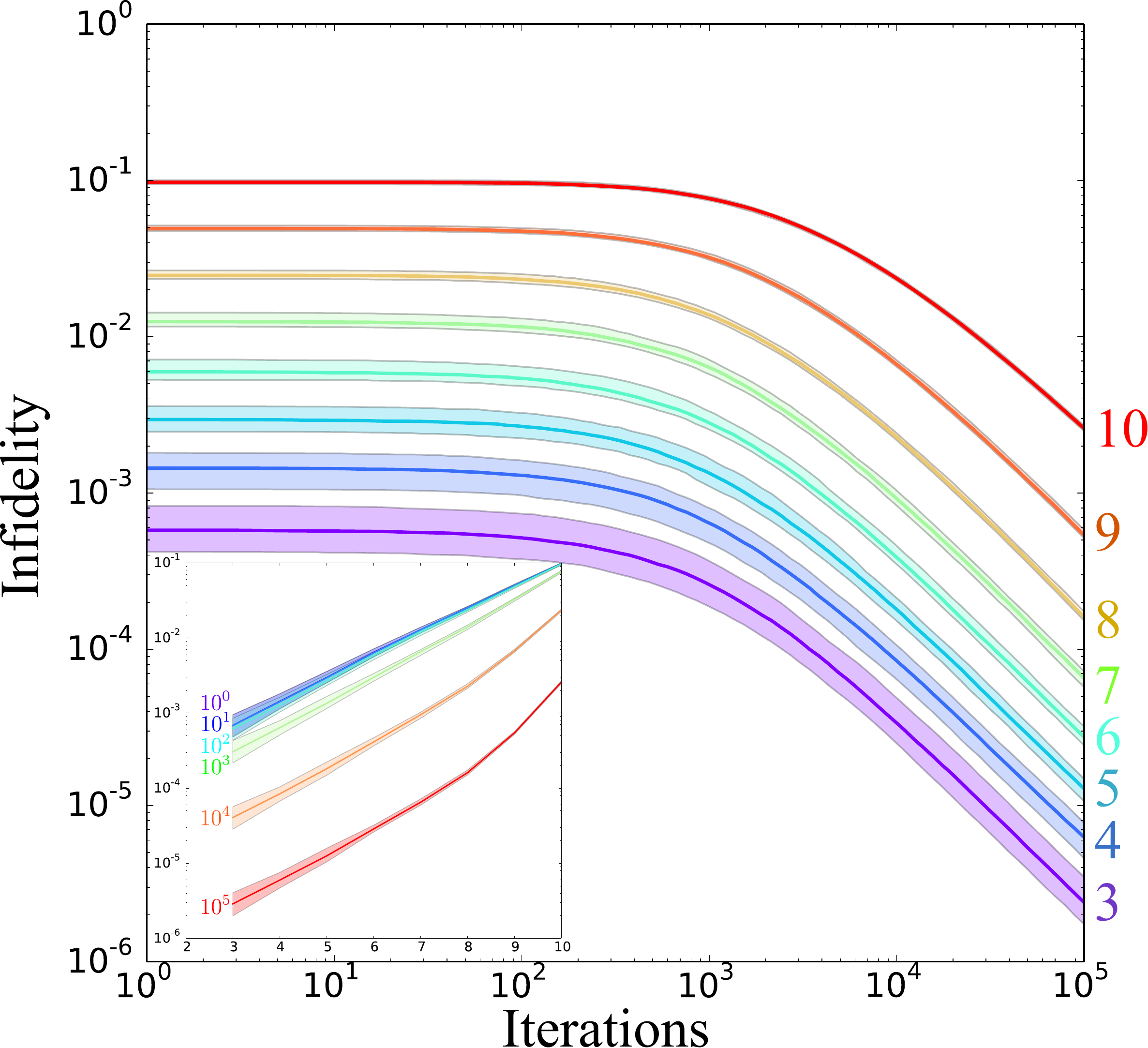}
  \caption{\label{fig:qubits_v_k} The infidelity vs the number of iterations $k$ achieved by SGQT for states of increasing qubit number (shown on the right).  The number of experiments per fidelity estimate for all cases was $N=10^4$.   Each line is the median performance of SGQT over 100 randomly (according to Haar measure) chosen pure states, where the initial guess was a random perturbation of 0.01 standard deviation in each dimension of the parametrized space (this was done to more efficiently extract the asymptotic scaling).  The shaded regions represent the interquartile range of infidelities.  The inset shows the performance as a function of the number of qubits for fixed values of $k$ as marked.  The SGQT parameters used for these simulations (and all further simulations) were $a=0.3$, $A=1000$ and $b=0.1$.}
\end{figure}

Above we have explored the efficacy of SGQT for single qubit tomography.  In Fig.~\ref{fig:qubits_v_k}, we generalize to multiple qubits.  The fits to the $O(1/k^\gamma)$ scaling gave values $\gamma\in(0.80,1.05)$, although it is difficult to trust this as an asymptotic fit for the data on $10$ qubits (which gave $\gamma=0.80$).  Separately fixing $k$ and fitting the infidelity to $O(d^\eta)$ (shown in the inset of Fig.~\ref{fig:qubits_v_k}) gave values $\eta\in(1.02,1.35)$.  As expected, since even estimating the fidelity to an arbitrary single pure target state is not efficient in the number of qubits, $n$, the convergence of SGQT is not efficient in the number of qubits.  

\begin{figure}\centering
 \includegraphics[width=1\columnwidth]{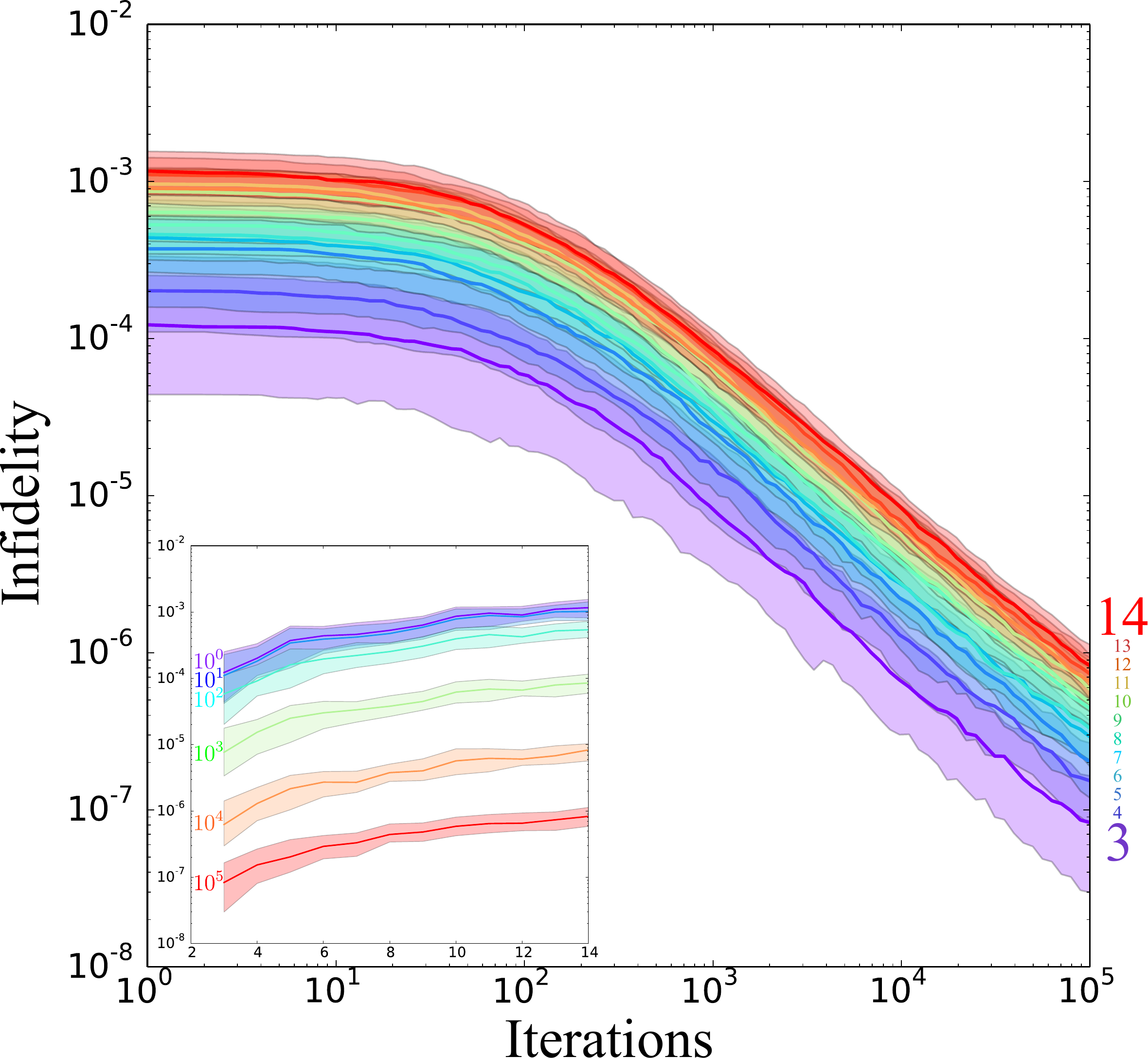}
  \caption{\label{fig:Wdepol} The rescaled infidelity (difference between the minimum achievable infidelity and actual infidelity) vs the number of iterations $k$ achieved by SGQT for W-states of increasing qubit number  (shown on the right). Parameters are as in Fig. \ref{fig:qubits_v_k}.}
\end{figure}

If, on the other hand, we restrict the class of states to one which can be efficiently specified and the fidelity to which can be efficiently estimated, SGQT becomes efficient.  As an example, consider the W-class of states which have found use in the theory of entanglement \cite{somma2006lower}.  An $n$ qubit W-class state is one of the form
\begin{equation}
\ket\psi = \alpha_1 \ket{10\ldots 0} + \alpha_2 \ket{01\ldots0} + \cdots +\alpha_n\ket{00\ldots1}.
\end{equation}
Note that the number of parameters grows linearly with the number of qubits.  Moreover, the fidelity to a target in this class can be estimated efficiently \cite{flammia2011direct, da2011practical}.  But if we actually want to \emph{learn} the state we are faced with an new problem: the actual true state might not lie in this subclass.  Using SGQT with fidelity estimation we can, however, efficiently find the W-class state with highest fidelity to the true state.  This points to both the robustness of SGQT and efficacy of solving the problem of finding the ``closest'' state within a desirable subclass.  The performance of SGQT for this problem is demonstrated in Fig.~\ref{fig:Wdepol}, where SGQT is shown to find the highest fidelity W-class state to a randomly generated mixed state.  The mixed state is generated by first Haar randomly choosing a W-state, then subjecting it to 5\% depolarizing noise.  Fits to $O(1/k^\gamma)$ asymptotic scaling for each qubit number lie in $\gamma\in(0.95,1.03)$, in line with the optimal performance.   Separately fixing $k$ and fitting the infidelity to $O(d^\eta)$ (shown in the inset of Fig.~\ref{fig:Wdepol}) gave values $\eta\in(1.41,1.53)$.  Since $d= 2(n-1)$, this is efficient in $n$.

\begin{figure}\centering
 \includegraphics[width=1\columnwidth]{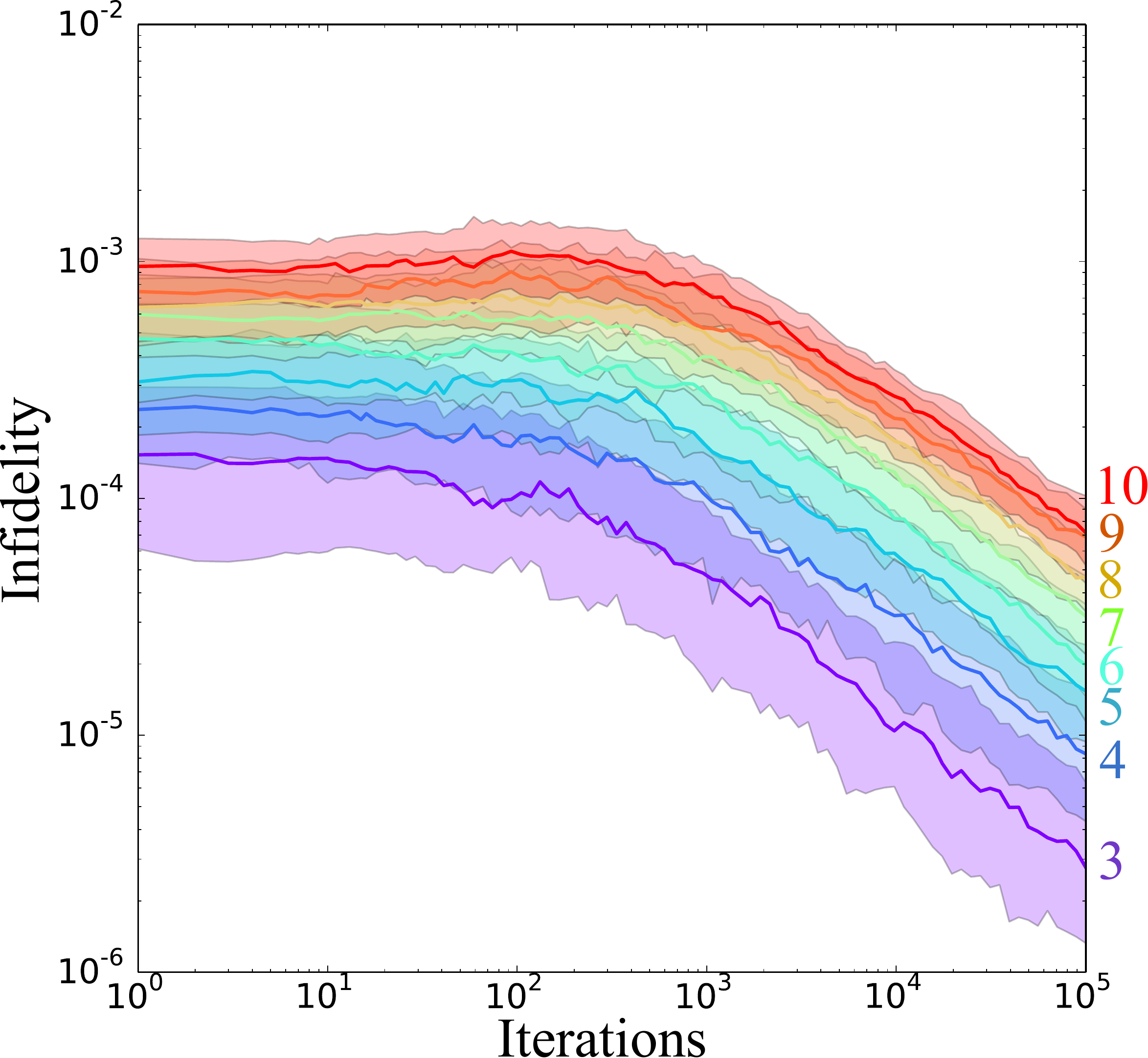}
  \caption{\label{fig:robust} The infidelity vs the number of iterations $k$ achieved by SGQT for W-states of increasing qubit number.  For each fidelity estimation measurement, an unknown zero-mean, 0.1 standard deviation, Gaussian random perturbation was applied to the target state.  All other parameters are as in Fig.~\ref{fig:qubits_v_k}.}
\end{figure}

Finally, we show that SGQT is robust to small amounts of measurement errors, which is a part of the ever-present \emph{state preparation and measurement} (SPAM) error problem.  In Fig.~\ref{fig:robust}, we demonstrate the robustness of SGQT to measurement errors for W-state estimation, where the measurement error is simulated by randomly perturbing the measurement target state with zero-mean Gaussian noise with a (quite high) standard deviation of 0.1.  The convergence is slower than noiseless measurements with fits giving $\gamma\in (0.48,0.59)$.  However, the convergence demonstrates that SGQT is robust to both statistical and technical measurement noise.

So SGQT works, but in what sense is it more efficient than standard quantum tomography (SQT)?  There are three things to consider: (1) how the infidelity scales with the total number of copies (here $N_{\text{tot}}$) of the system and the dimension ($d$); (2) the total number of different measurement directions; (3) the space complexity (number of real numbers to store); and (4) the computational time required to arrive at the final estimate.  

Asymptotic arguments relate fidelity to Euclidean distance on some parameterization of pure states which transform the problem into one with a known solution from classical statistics \cite{class}.  The infidelity of estimating pure states is at least $O(d/N_{\text{tot}})$ and can be achieved by maximum likelihood estimation SQT.  The extracted scaling of SGQT is slightly worse giving $O(d^\eta/N_{\rm tot})$ with $\eta>1$.  Noting that the implementation of the SGQT algorithm was written with a single line of code and a static set of algorithmic parameters, we think it is quite good for a first investigation in self-learning algorithms.  With more sophisticated numerical algorithms, we conjecture that self-learning approaches can achieve $O(d/N_{\text{tot}})$ and with hopefully less effort than has put into the theoretical analysis of tomography.

The total number of measurement directions for SGQT is $O(N_{\text{tot}})$, whereas for tomography it is a free parameter $m_{\text{tot}}$.  For informationally complete tomography $m_{\text{tot}} =O(d)$, but overcomplete measurement sets are often considered.  Finally, where SGQT can provide substantial improvements in efficiency is in the computational complexity of calculating the estimator.  The amount of storage required for tomography is at least $O\left({m_{\text{tot}}}d\right)$, the number of elements in a measurement consisting of rank-1 elements times the space required to store each element.  Since SGQT is online, it forgets the past measurements and only requires $O\left(d\right)$  storage space.  To actually compute an estimator from a data set requires at least  $O\left({m_{\text{tot}}}d^2\right)$ time (the complexity of linear least squares regression), with more sophisticated techniques (such as maximum likelihood) taking much longer.  Since SGQT ends with an estimate of the state, there is no computation needed---an enormous advantage.  These considerations are summarized in Table \ref{table}.

\begin{table}
\begin{tabular}{lcccc} 
& Infidelity & Measurements & Space & Time\\
SQT & $O\left( d/ {N_{\text{tot}}}\right)$ & {\color{green} $m_{\text{tot}}$} & {\color{red} $O\left({m_{\text{tot}}}d\right)$} & {\color{red} $O\left({m_{\text{tot}}}d^2\right)$}\\
SGQT & $O\left(d^\eta /{N_{\text{tot}}}\right)$ & {\color{red} $O\left({N_{\text{tot}}}\right)$} & {\color{green} $O\left(d\right)$} & {\color{green} $O\left(d\right)$}\\
\end{tabular}\caption{\label{table} Summary of the trade-off in complexities for SGQT and SQT  (standard quantum tomography).  Numerical fits here give $\eta>1$, but it is conjectured that $\eta = 1$ is achievable with optimized choices on gain sequences $\{\alpha_k\}$ and $\{\beta_k\}$. }
\end{table}

Here we have considered examples of pure state tomography using infidelity since it has a clear-cut interpretation and is a standard error metric.  We reiterate that SGQT will work with \emph{any} distance metric so long as it is estimable via experiment.  The only caveat is that the efficiency of SGQT is directly related to the efficiency with which the distance can be estimated.  For example, in direct fidelity estimation \cite{flammia2011direct, da2011practical}, only certain classes of states can be validated in an efficient way, such as our W-state example.  If only a subclass of states is considered, SGQT will converge to the nearest state within that subclass, as we have demonstrated with W-states.  We have also shown that SGQT is robust to certain forms of SPAM errors.  In the same way as for states, SGQT can be used to find quantum channels where randomized benchmarking \cite{moussa2012practical, magesan2012efficient} can be used to efficiently estimate the fidelity to certain classes of unitaries.
Finally, we note that to further mitigate the issues of complexity, it may become viable in the future to aid the estimation of the distance measure with quantum resources \cite{wiebe2013hamiltonian, wiebe_2014_quantum}, such as the swap test \cite{buhrman2001quantum}.

In summary, we have provided an experimental protocol to learn quantum states without the need for classical reconstruction---that is, the quantum system guides itself to a description of its own state.  Using ideas from stochastic optimization theory and the direct estimation of fidelity, we have shown that certain classes of states can be learned efficiently in an iterative experimental protocol which ends with the experiment determining its own state.  This result demonstrates that the standard, and prohibitive, paradigm of first collecting massive amounts of data, then solving the inverse problem of state estimation is unnecessary.  Perhaps with an eye to the future, the approach considered here is a step toward quantum learning, in which autonomous quantum machines learn and manipulate their environment without the need of a human operator. 

\begin{acknowledgements}
The author thanks Robin Blume-Kohout for helpful discussions.  This work was supported in part by National Science Foundation Grant No. PHY-1212445 and by the Canadian Government through the NSERC PDF program.
\end{acknowledgements}

\end{document}